\documentclass[aps,prd,twocolumn,superscriptaddress,amsmath,amssymb,amsthm,nofootinbib, preprintnumbers]{revtex4-2}
\usepackage{url}
\usepackage[colorlinks=true,
    linkcolor=blue,
    citecolor=blue,
    urlcolor=blue]{hyperref}
 \usepackage{amsmath}
\usepackage{graphicx}
\usepackage{epstopdf}
\usepackage{float}
\usepackage{hyperref}
\usepackage{color}
\usepackage[T1]{fontenc}
\usepackage[utf8]{inputenc}
\usepackage[toc,page]{appendix}
\usepackage[usenames,dvipsnames]{xcolor}
\usepackage[normalem]{ulem}

\usepackage{tikz}
\usetikzlibrary{shapes,positioning,shadows,graphs.standard,automata,arrows}
\usepackage{pgfplots}
\pgfplotsset{compat=newest, width=2.669cm, height=2.669cm, scale only axis=true,enlargelimits=false}
\pgfplotsset{tick label style={font=\tiny}}
\pgfplotsset{every major tick/.append style={major tick length=3pt}}
\pgfplotsset{every minor tick/.append style={minor tick length=1.5pt}}
\usepgfplotslibrary{groupplots}
\usepackage{caption}
\usepackage[font=small,labelfont=bf,justification=justified,format=plain]{caption}

\providecommand{\renewoperator}[3]{%
\renewcommand*{#1}{\mathop{#2}#3}}

\makeatletter
\providecommand*{\diff}%
{\@ifnextchar^{\DIfF}{\DIfF^{}}}
\def\DIfF^#1{%
\mathop{\mathrm{\mathstrut d}}%
\nolimits^{#1}\gobblespace}
\def\gobblespace{%
\futurelet\diffarg\opspace}
\def\opspace{%
\let\DiffSpace\!%
\ifx\diffarg(%
\let\DiffSpace\relax
\else
\ifx\diffarg[%
\let\DiffSpace\relax
\else
\ifx\diffarg\{%
\let\DiffSpace\relax
\fi\fi\fi\DiffSpace}

\providecommand*{\eu}{\ensuremath{\mathrm{e}}}
\providecommand*{\iu}{\ensuremath{\mathrm{i}}}

\renewoperator{\Re}{\mathrm{Re}}{\nolimits}
\renewoperator{\Im}{\mathrm{Im}}{\nolimits}

\usepackage{bm}

\newcommand{\be}{\begin{equation}}
\newcommand{\ee}{\end{equation}}
\newcommand{\ba}{\begin{eqnarray}}
\newcommand{\ea}{\end{eqnarray}}

\newcommand{\beq}{\begin{equation}}
\newcommand{\eeq}{\end{equation}}
\newcommand{\beqa}{\begin{eqnarray}}
\newcommand{\eeqa}{\end{eqnarray}}

\begin{document}

\title{A new exact rotating spacetime in vacuum: The Kerr--Levi-Civita Spacetime}

\author{Jos{\'e} Barrientos}
\email{jbarrientos@academicos.uta.cl}
\affiliation{Sede Esmeralda, Universidad de Tarapac{\'a}, Avenida Luis Emilio Recabarren 2477, Iquique, Chile}
\affiliation{Institute of Mathematics of the Czech Academy of Sciences, {\v Z}itn{\'a} 25, 115 67 Prague 1, Czech Republic}
\affiliation{Vicerrector\'ia de Investigaci\'on y Postgrado, Universidad de La Serena, La Serena 1700000, Chile}

\author{Adolfo Cisterna}
\email{adolfo.cisterna@mff.cuni.cz}
\affiliation{Sede Esmeralda, Universidad de Tarapac{\'a}, Avenida Luis Emilio Recabarren 2477, Iquique, Chile}
\affiliation{Institute of Theoretical Physics, Faculty of Mathematics and Physics,
Charles University, V Hole{\v s}ovick\'ach 2, 180 00 Prague 8, Czech Republic}

\author{Mokhtar Hassaine}
\email{hassaine@inst-mat.utalca.cl}
\affiliation{Instituto de Matem\'{a}ticas,
Universidad de Talca, Casilla 747, Talca, Chile}

\author{Keanu M{\"u}ller}
\email{keanumuller2016@udec.cl}
\affiliation{Departamento de F\'isica, Universidad de Concepci\'on,
Casilla, 160-C, Concepci\'on, Chile}

\author{Konstantinos Pallikaris}
\email{konstantinos.pallikaris@ut.ee}
\affiliation{Laboratory of Theoretical Physics, Institute of Physics, University of Tartu, W. Ostwaldi 1, 50411 Tartu, Estonia}


\begin{abstract}
We construct a new rotating solution of Einstein's theory in vacuum by exploiting the Lie point symmetries of the field equations in the complex potential formalism of Ernst. 
In particular, we perform a discrete symmetry transformation, known as inversion, of the gravitational potential associated with the Kerr metric. The resulting metric describes a rotating generalization of the Schwarzschild--Levi-Civita spacetime, and we refer to it as the Kerr--Levi-Civita metric. We study the key geometric features of this novel spacetime, which turns out to be free of curvature singularities, topological defects, and closed timelike curves. These attractive properties are also common to the extremal black hole and the super-spinning case. The solution is algebraically general (Petrov-type I), and its horizons lie at the horizon radii of the Kerr black hole. 
The ergoregions, however, are strongly influenced by the Levi-Civita-like asymptotic structure, producing an effect akin to the magnetized Kerr--Newman and swirling solutions.
Interestingly, while its static counterpart permits a Kerr--Schild representation, the Kerr--Levi-Civita metric does not admit such a formulation.

\end{abstract}

\maketitle

\section{Introduction}
To date, only two exact solutions with well-defined static limits are known to describe the exterior gravitational field of a spinning mass in vacuum. The first is the Kerr metric \cite{Kerr:1963ud}, which remains the most astrophysically relevant solution to Einstein’s field equations, underpinning much of our theoretical understanding of rotating black holes. The second is its less familiar generalization, the Tomimatsu--Sato metric \cite{Tomimatsu:1972zz}. Unlike the Kerr metric, the Tomimatsu--Sato solution does not reduce to the Schwarzschild geometry in the limit of vanishing angular momentum, but to a class of Weyl metrics known as Zipoy--Voorhees spacetimes \cite{Zipoy:1966btu,Voorhees:1970ywo}.

The circular, stationary, and axisymmetric Einstein and Einstein--Maxwell field equations become remarkably integrable in the complex-potential formalism of Ernst~\cite{Ernst:1967by,Ernst:1967wx}. That formalism reveals a hidden set of Lie point symmetries which, under appropriate conditions, serves as a powerful tool for generating sophisticated solutions \cite{Kinners,Geroch:1970nt,Ehlers:1959aug,Harrison}. 

In this letter, we construct a novel rotating solution to Einstein’s field equations in vacuum by ``inverting'' the gravitational potential associated with the Kerr metric in magnetic Weyl--Lewis--Papapetrou (WLP) form. The new metric, dubbed \emph{Kerr--Levi-Civita} (Kerr--LC) metric, is also a solution to the field equations, because inversion is a discrete symmetry of the Ernst equations. In the static limit, it reduces to a metric describing a Schwarzschild black hole embedded into a Levi-Civita (LC) cylindrical background~\cite{Levi-Civita}.

The Kerr--LC spacetime features a notable amount of attractive properties: it is entirely free of curvature singularities, conical defects, spinning strings, and closed timelike curves. Such properties hold in all regimes (non-extremal, extremal, and super-spinning). According to the Petrov classification \cite{Stephani:2003tm}, the spacetime is algebraically general. Its inner and outer horizons are located at the radii of the respective horizons of the Kerr black hole, while the ergoregion has a frame-dependent profile quite different from the usual Kerr ergosphere, rather reminiscent of the respective profiles in the magnetized Kerr--Newman spacetime~\cite{Gibbons:2013yq} and in swirling black holes. Although its static limit, the Schwarzschild--LC black hole, admits a Kerr--Schild representation \cite{Barrientos:2024uuq}, the Kerr--Levi-Civita metric does not.

This letter is organized as follows. In Sec.~\ref{secII}, we present a detailed construction of the solution. In Sec.~\ref{secIII}, we study the key geometric properties of the Kerr--LC spacetime. Finally,  we conclude in Sec.~\ref{secIV}, discussing open questions and potential directions for future research.
\vspace{-0.5cm}
\section{Construction of the Kerr--Levi-Civita spacetime}\label{secII}
Stationary and axisymmetric geometries in four dimensions are characterized by the action of a group $\mathbb{R} \times \mathbf{SO(2)}$, under which the spacetime metric remains invariant. If the surfaces of transitivity of that group (its orbits) are everywhere orthogonal to a family of hypersurfaces defined by the (remaining) non-Killing coordinates (known as meridional surfaces), the spacetime is said to be circular, and the action of $\mathbb{R} \times \mathbf{SO(2)}$ is said to be orthogonally transitive \cite{Carter:1969zz,CarterRe}.\footnote{Note that, in the case of stationary, axisymmetric spacetimes in Einstein--Maxwell theory, the circularity condition is not an assumption, but rather a consequence of the field equations, provided that the electromagnetic field-strength tensor is compatible with the symmetry ansatz~\cite{CarterRe}.}

The most general class of circular, stationary, and axisymmetric spacetimes is described by the WLP metric, either in its standard electric form or in its magnetic representation. Note that the latter is obtained by means of a double Wick rotation that transforms the temporal and azimuthal coordinates. In the canonical Weyl coordinates denoted by $\{t,\rho,z,\phi\}$, the magnetic WLP line element reads 
\begin{equation}
\diff s^2 = f \left( \diff \phi - \omega\diff t \right)^2 + \frac{1}{f} \left[ \eu^{2\gamma} \left(\diff \rho^2 + \diff z^2\right) - \rho^2 \diff t^2 \right], \label{WLPm}
\end{equation}
where $f,\ \omega$, and $\gamma$ are functions of $\rho$ and $z$ only.

It is well established that for circular, stationary, and axisymmetric spacetimes expressed in a canonical WLP form, Einstein’s equations can be recast in the Ernst formulation as the complex Ernst equation \cite{Ernst:1967by,Ernst:1967wx}
\begin{align}
\operatorname{Re} \left(\mathcal{E}\right) \nabla^2 \mathcal{E}=\bm{\nabla} \mathcal{E} \cdot\bm{\nabla} \mathcal{E}.
\end{align}
Here, the complex Ernst gravitational potential
\begin{align}
\mathcal{E}=-f-\iu \chi ,
\end{align}
is determined by the characteristic metric functions $f$ and $\omega$ in the magnetic WLP line element~\eqref{WLPm} and the relevant twist equation
\begin{align}
\bm{\hat{\phi}} \times \bm\nabla \chi & =-\frac{f^2}{\rho} \bm\nabla \omega, \label{magtwist}
\end{align}
that provides the twisted potential $\chi$. It is worth mentioning that the Laplacian and the gradient are defined with respect to the flat 3-metric in cylindrical coordinates $\{\rho, z, \phi\}$.

The main advantage of this sophisticated formulation
of Einstein's equations lies in the revelation of
a set of otherwise hidden Lie point symmetries intrinsic
to the Ernst equations \cite{Kramer}. Relevant to our construction is the inversion transformation, a discrete symmetry transformation in potential space, obtained via the composition of a duality-rescaling transformation, a gravitational gauge transformation, and an Ehlers transformation~\cite{Ehlers:1959aug} (see Sec.~\ref{secIV} for the details).\footnote{In~\cite{Stephani:2003tm}, the authors present inversion as a composition of the above transformations in the limit $j\to 0$. In Sec.~\ref{secIV}, we make it clear why taking the limit is a redundant action.} 

Denoting seed quantities with a $0$ subscript, the action of the inversion operator on the gravitational Ernst potential $\mathcal{E}_0=-f_0-\iu\chi_0$ associated with a vacuum seed, is defined by 
\begin{equation}
\label{inversion}
\mathrm{I}:\mathcal{E}_0 \mapsto\mathcal{E}:=\frac{1}{\mathcal{E}_0}.
\end{equation}
Hence, the transformed Ernst potential $\mathcal{E}=-f-\iu\chi$ is explicitly given by 
\begin{equation}
    {f}=\frac{f_0}{f_0^2+\chi_0^2},\qquad {\chi}=-\frac{\chi_0}{f_0^2+\chi_0^2}. \label{functionsnew}
\end{equation}
Since the function $\gamma$ remains invariant under the Lie point symmetries, and can always be expressed in terms of $\mathcal{E}$ using a pair of quadratures \cite{Stephani:2003tm}, the ``novelty'' of the target spacetime is entirely encoded in the functions $f$ and the ``rotation function'' $\omega$, which is obtained from the twist equation \eqref{magtwist} once $\chi$ has been determined.

To construct our Kerr--Levi-Civita spacetime, we begin with the Kerr metric as the seed, written in magnetic WLP form using spherical-like coordinates $\{t,r,x=\cos\theta,\phi\}$ for simplicity. Explicitly, we have that
    \begin{equation}
    \diff s^2_0\!=\!f_0(\diff\phi - \omega_0\diff t)^2 - \frac{\Delta_r\Delta_x\diff t^2}{f_0}
    +\frac{\eu^{2\gamma_0}}{f_0}\left(\frac{\diff r^2}{\Delta_r}+\frac{\diff x^2}{\Delta_x}\right) ,\label{mWLPspherical}
\end{equation}
where
\begin{equation}
\begin{aligned}
f_0(r,x)&=\Delta_x\frac{(r^2+a^2)^2-a^2\Delta_r\Delta_x}{\varrho^2},\\
\omega_0(r,x)&=a\frac{(r^2+a^2)-\Delta_r}{(r^2+a^2)^2-a^2\Delta_r\Delta_x},\\
\eu^{2\gamma_0(r,x)}&=\Delta_x\left[(r^2+a^2)^2-a^2\Delta_r\Delta_x\right],\\
\Delta_r(r)&=r^2-2mr+a^2,\quad \Delta_x(x)=1-x^2,
\end{aligned}
\end{equation}
and $\varrho^2(r,x) = r^2 + a^2 x^2$.

The new metric function $f$ is readily obtained. The rotation function $\omega$ follows from the following procedure: starting from the known seed function $\omega_0$, we apply the twist equation \eqref{magtwist} to compute the imaginary part of the seed Ernst potential, $\chi_0$. This is then used in \eqref{functionsnew} to determine the imaginary component $\chi$ of the transformed potential. Substituting back into the twist equation \eqref{magtwist}, now with the transformed quantities, yields the function $\omega$. The gradient should now be understood as that defined with respect to the flat metric in the spherical-like coordinates, using the relations
\begin{equation}
    \rho(r,x)=\sqrt{\Delta_r\Delta_x},\quad z(r,x)=(r-m)x.
\end{equation}
\begin{widetext}
Thus, we end up with
\begin{align}
{f}(r,x) &=\Delta_x\varrho^2\frac{(r^2+a^2)^2-a^2\Delta_r\Delta_x}{4m^2a^2x^2\left[a^2\Delta_x^2+\varrho^2(\Delta_x+2)\right]^2+\Delta_x^2\left[(r^2+a^2)^2-a^2\Delta_r\Delta_x\right]^2},\\
{\omega}(r,x) &= -2 m a\frac{ 
        (2 a^2 m - 3 a^2 r + r^3)\Delta_r x^4
        - 6 r (a^2 + r^2)\Delta_r x^2
        + (2 a^2 m + a^2 r + r^3)(a^2 - 6 m r - 3 r^2)
    }{\Delta_r a^2 x^2 + r (2 a^2 m + a^2 r + r^3)}. 
\end{align}
\end{widetext} 
The complete Kerr--Levi-Civita spacetime is thus described in spherical-like coordinates by the line element \eqref{mWLPspherical} where $f_0=f$ and $\omega_0=\omega$ ($\gamma_0$ kept as is).
\vspace{-0.4cm}
\section{The Kerr--LC geometry\label{secIII}}
We now explore the features of this new geometry. Let us begin by observing that $g_{\phi\phi}\equiv f$ and that therefore the azimuthal metric component is always positive. It follows that the Killing vector $\bm{\partial}_\phi$ is everywhere spacelike for $r>0$, except on the symmetry axis, $x=\pm1$, where it becomes null. Consequently, there are no closed timelike curves. Second, let us introduce a new coordinate system $\{t, \tilde\rho, \tilde{z}, \phi\}$, where
\begin{equation}
    \tilde\rho := \frac{4ma}{|x|}\sqrt{a^2 + r^2 x^2}\sqrt{1 - x^2},\quad \tilde{z} := r x. \label{eq:ConicalSingularityCoords}
\end{equation}
In the latter, the induced metric on the slices of constant $t$ and $\tilde{z}$,  near the symmetry axis, becomes
\begin{equation}
  \diff s^2 \sim \diff {\tilde\rho}^2 + \frac{\tilde{\rho}^2}{256 a^4 m^4}\diff\phi^2.
\end{equation}
Since $|\bm{\partial}_\phi|^2 \sim \tilde\rho^2$, there is no cosmic spinning string there.\footnote{The absence of a Misner string follows.} However, there is a defect angle $2\pi(1-\vartheta)$ where $\vartheta=\pm 1/(16a^2m^2)$. To deal with that, we can redefine 
\begin{equation}
    \phi = 16 a^2 m^2 \varphi,
\end{equation}
and take $\varphi$ as our new azimuthal coordinate with $\varphi$ and $\varphi+2\pi$ identified. The new spacetime is free of conical singularities as well. 

In the coordinate system $\{t,\rho,z,\varphi\}$, the asymptotic form of the metric reads 
\begin{equation}
  \diff s^2 \sim \diff s^2_{\mathrm{LC}} - 64a^3m^3 \frac{3\rho^4 + 12 \rho^2 z^2 + 8 z^4}{\rho^2 (\rho^2 + z^2)^{3/2}}\, \diff t \diff \varphi,\label{eq:AsyForm}
\end{equation}
where
\begin{equation}
  \diff s^2_{\mathrm{LC}} = \rho^4 \left( -\diff t^2 + \diff \rho^2 + \diff z^2 \right) + \frac{256 a^4 m^4}{\rho^2}\diff \varphi^2 ,\label{eq:AlternativeLCmetric}
\end{equation}
is an alternative formulation of the Levi-Civita metric with $\sigma=1=B$ and $C = 16 a^2 m^2$~\cite{Griffiths:2009dfa}. To get equation~\eqref{eq:AsyForm} we took the limit $\rho,z\to\infty$ while keeping $\rho/z$ finite.\footnote{Amounts to $r\to\infty$ in Boyer-Lindquist coordinates.} Since the asymptotic geometry is a rotating generalization of the LC geometry, we may formally address the solution as the Kerr--LC spacetime. 

We shall now attack the horizon structure. The metric component $g_{rr}$ features poles at the outer and inner horizon radii of the Kerr black hole,
\begin{equation}
    r_+ = m + \sqrt{m^2 - a^2},\quad r_- = m - \sqrt{m^2 - a^2},
\end{equation}
respectively. In more detail, we have that $g_{rr}\sim 1/\Delta_r$ as $r\to r_\pm$. Nevertheless, curvature invariants are regular there. Surprisingly, they are regular everywhere (see App.~\ref{appA}); there is no \emph{ring singularity} (like in Kerr). Moreover, tidal forces either vanish or are $\mathcal{O}(1)$ in all directions at infinity, and we may conclude that the Kerr--LC metric, which is a Petrov-type I solution to Einstein's field equations, describes a regular black hole for $m^2\geq a^2$. On the other hand, for $m^2<a^2$, the geometry is regular and free of horizons. 

\begin{figure}
        \centering
        \begin{tikzpicture}
        \begin{groupplot}[group style={group size=2 by 2, horizontal sep=0.51cm, vertical sep=0.51cm}]
        \nextgroupplot[%
        enlargelimits=false,
        axis on top,
        xlabel= $X$,
        ylabel= $Z$,
        xlabel near ticks,
        ylabel near ticks,
        xticklabel pos=top,
        yticklabel pos=left,
        scaled y ticks=false,
        scaled x ticks=false,
        minor x tick num=5,
        minor y tick num=5,
        ytick distance = {4},
        xtick distance = {4}
        ]
        \draw[lightgray] (axis cs:-8,0) -- (axis cs:8,0);
        \draw[lightgray] (axis cs:0,-8) -- (axis cs:0,8);
        \addplot graphics
            [xmin=-8,xmax=8,ymin=-8,ymax=8]
                {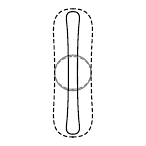};
        \node[anchor= center] at (rel axis cs:.1,.9) 
        {(a)};
        \nextgroupplot[%
        enlargelimits=false,
        axis on top,
        xlabel= $X$,
        ylabel= $Z$,
        xlabel near ticks,
        ylabel near ticks,
        xticklabel pos=top,
        yticklabel pos=right,
        scaled y ticks=false,
        scaled x ticks=false,
        minor x tick num=5,
        minor y tick num=5,
        ytick distance = {4},
        xtick distance = {4}
        ]
        \draw[lightgray] (axis cs:-8,0) -- (axis cs:8,0);
        \draw[lightgray] (axis cs:0,-8) -- (axis cs:0,8);
        \addplot graphics
            [xmin=-8,xmax=8,ymin=-8,ymax=8]
                {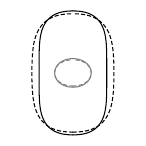};
        \node[anchor= center] at (rel axis cs:0.1,.9) 
        {(b)};
        \nextgroupplot[%
        enlargelimits=false,
        axis on top,
        xlabel= $X$,
        ylabel= $Z$,
        xlabel near ticks,
        ylabel near ticks,
        xticklabel pos=bottom,
        yticklabel pos=left,
        scaled y ticks=false,
        scaled x ticks=false,
        minor x tick num=5,
        minor y tick num=5,
        ytick distance = {4},
        xtick distance = {4}
        ]
        \draw[lightgray] (axis cs:-8,0) -- (axis cs:8,0);
        \draw[lightgray] (axis cs:0,-8) -- (axis cs:0,8);
        \addplot graphics
            [xmin=-8,xmax=8,ymin=-8,ymax=8]
                {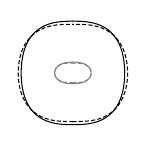};
        \node[anchor= center] at (rel axis cs:0.1,.9) 
        {(c)};
        \nextgroupplot[%
        enlargelimits=false,
        axis on top,
        xlabel= $a$,
        ylabel= ${K}$,
        xlabel near ticks,
        ylabel near ticks,
        xticklabel pos=bottom,
        yticklabel pos=right,
        scaled y ticks=false,
        scaled x ticks=false,
        minor y tick num=1,
        minor x tick num=1,
        ytick distance = {0.1},
        xtick distance = {0.2}
        ]
        \draw[lightgray] (axis cs:0.2,0) -- (axis cs:1,0);
        \addplot graphics
            [xmin=0.2,xmax=1,ymin=-0.1,ymax=0.4]
                {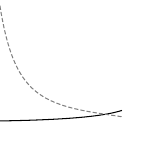};
        \node[anchor= center] at (rel axis cs:0.1,.9) 
        {(d)};
        \end{groupplot}
        \end{tikzpicture}
        \caption[]{In panels (a), (b), and (c), we show a cross section, taken at $Y=0$, of the embeddings of the outer-horizon surfaces of Kerr (gray) and Kerr--LC (black) black holes for $a=1/4$ (solid) and $a=1/2$ (dashed), $a=a_0$ (solid) and $a=0.75$, and $a=a_{\mathrm{iso}}\approx 0.84$ (solid) and $a=a_{\max}$ (dashed), respectively. The function $Z$ is the embedding function and $X^2+Y^2 = \mathcal{R}^2$, where $\mathcal{R}$ is the proper circumferential radius at $r=r_{\!+}$. In panel (d), we graph the Gaussian curvature ${K}$ of the outer horizon at the equator (black, solid) and the poles (gray, dashed). The solid curve crosses ${K}=0$ exactly at $a=a_0$. The two curves meet at $a=a_{\mathrm{iso}}$ and terminate at $a=a_{\max}$. We have set $m=1$.}
        \label{fig:Horizon}
        \end{figure}

Of course, we expect that the horizon surface will not be the same as that of the Kerr black hole. For example, the equatorial circumference of the outer horizon\textemdash{}independent of the spin parameter (or specific angular momentum) $a$ in Kerr spacetimes and equal to $4\pi m$\textemdash{}reads $C_{\mathrm{eq}}=16\pi m a^2$ in our case. Thus, for a Kerr--LC black hole, higher angular momentum results in an outer horizon with larger equatorial circumference for fixed mass. In \autoref{fig:Horizon}, we present a cross-section of the embeddings of the outer-horizon surfaces of Kerr and Kerr--LC black holes in three-dimensional Euclidean space for different values of the spin parameter.
We do so in order to visualize the deformation of the horizon in the new solution. Embeddings are valid for $|a|<a_{\max} = \sqrt{3}m/2$. For $a<a_0 \approx 0.68 m$, we have a peanut-shaped horizon with the neck located at the equator. At $a=a_0$, the Gaussian curvature at the equator is 0, and for $a_0<a<a_{\mathrm{iso}}$, the surface becomes a prolate capped cylinder. For $a=a_{\mathrm{iso}}$, the horizon is an isocircumferential barrel, that becomes oblate for $a_{\mathrm{iso}}<a<a_{\max}$. 

Note also that the horizon $r=r_+$ is a Killing horizon. The linear combination $\bm{\partial}_t + \Omega|_{\mathrm{H}}\bm{\partial}_\varphi$, where 
\begin{equation}
    \Omega|_{\mathrm{H}} = \frac{3r_+ + r_-}{2\sqrt{r_+ r_-}(r_++r_-)}\label{eq:BHAV},
\end{equation}
is the angular velocity of the black hole, is a Killing vector that becomes null on the event-horizon surface. Recall that the angular velocity of the black hole is the angular velocity of a zero angular momentum observer (ZAMO)\footnote{The ZAMO angular velocity reads $\Omega = u^\varphi/u^t$ where $\bm{u}$ is the four-velocity of the observer satisfying $\bm{u}\cdot \bm{\partial}_\varphi=0$.} at $r=r_+$, which in our case, takes the value~\eqref{eq:BHAV} on the subtle assumption that $\bm{\partial}_t$ is the asymptotically timelike Killing vector. However, as pointed out in~\cite{Barrientos:2024pkt}, the concept of ``rest'' is ambiguous in spacetimes with an asymptotic behavior similar to that in equation~\eqref{eq:AsyForm}, see also Ref.~\cite{Gibbons:2013yq}. To be more precise, there is no Killing vector that is timelike \emph{everywhere} at infinity, and $\bm{\partial}_{t^{(\alpha)}}=\bm{\partial}_t+\alpha \bm{\partial}_\varphi$ all define a time flow in some parts of the asymptotic region. One should, in theory, make that ambiguity manifest in the spacetime metric by moving to a rotating frame, i.e., switching over to coordinates
\begin{equation}
\{t^{(\alpha)}=t,r,x,\varphi^{(\alpha)}=\varphi - \alpha t\},\label{eq:RotFrame}
\end{equation}
adapted to $\bm{\partial}_{t^{(\alpha)}}$. For our purposes, it suffices to consider
\begin{equation}
    |\bm{\partial}_{t^{(\alpha)}}|^2\geq 0,\quad \Omega^{(\alpha)}=-\frac{\bm{\partial}_{t^{(\alpha)}}\cdot \bm{\partial}_{\varphi^{(\alpha)}}}{|\bm{\partial}_\varphi|^2},
\end{equation}
as the definitions for the ergoregions and the angular velocity of a ZAMO, respectively, having in mind that fixing $\alpha$ amounts to choosing a frame. Note that $\Omega^{(\alpha)} = \Omega^{(0)}-\alpha$, becomes constant at $r=r_+$. We shall address that constant, call it $\Omega^{(\alpha)}|_{\mathrm{H}}$, as the \emph{relative} angular velocity of the black hole. Let us remark that velocity differences like $\Delta\Omega = \Omega^{(\alpha)}-\Omega^{(\alpha)}|_{\mathrm{H}}$, for example, do not depend on $\alpha$; they are absolute quantities in that regard. By construction, the Killing vector $\bm{\partial}_{t^{(\alpha)}}+\Omega^{(\alpha)}|_\mathrm{H} \bm{\partial}_{\varphi}$ is null on the event-horizon surface in all rotating frames. 

Some more comments on the angular velocity and the nature of the rotation in the Kerr--LC spacetime are in order. In a Kerr spacetime, the asymptotic geometry is static; the angular velocity falls off like $\sim 2am/r^3$ as $r\to\infty$. In the Kerr--LC spacetime, we have that 
\begin{equation}
    \Omega \sim \frac{3+ 6 x^2 - x^4}{8am}r,\label{eq:AsyAV}
\end{equation}
at large distances and fixed latitude. It is not surprising that the above expression does not contain $\alpha$. The reason is that the asymptotic geometry is insensitive to the choice of frame; all ZAMOs will experience the same dragging far away. That dragging, however, does not fit in the picture of an isolated, massive, rotating compact object (like the Kerr black hole) ``dragging along'' spacetime in its vicinity. It is rather reminiscent of what happens in spacetimes describing black holes embedded into a swirling universe~\cite{Contopoulos:2015wra,Barrientos:2024pkt,DiPinto:2024axv,DiPinto:2025yaa},\footnote{Recall that the swirling background is asymptotic to a rotating generalization of the LC metric with $\sigma=1/4$.} although asymptotic frame-dragging should be common to all solutions with rotating asymptotic forms. After all, equation~\eqref{eq:AsyAV} is the ZAMO angular velocity in the asymptotic spacetime~\eqref{eq:AsyForm}. However, there is a crucial difference between the Kerr--LC solution and, for example, the swirling black holes. 

The frame-dragging in the latter exists even if we remove the mass source, be it static or non-static, due to the intrinsic rotation of the background. On the other hand, dragging in the Kerr--LC spacetime is solely due to the angular momentum $J\equiv am$ of the Kerr seed. In the topologically regular solution, the limits $m,J\to0$ ($a$ finite) or $J\to 0$ ($a\to 0$ with $m$ fixed) result in a singular metric since $g_{\varphi\varphi}\to 0$. The price we pay for removing the conical singularity is that we can no longer remove the source or its angular momentum. For the sake of the argument, however, let us consider the spacetime with a conical defect and take the limits $m\to 0$ and $J\to 0$ while keeping $a$ finite. The resulting metric is
\begin{equation}
    \diff s^2 = \mathring{\rho}^4\left(-\diff t^2 +\diff\mathring{\rho}^2+\diff\mathring{z}^2\right) +\frac{\diff\phi^2}{\mathring{\rho}^2},\label{eq:LClimit}
\end{equation}
where $\mathring{\rho}$ and $\mathring{z}$ are the values of $\rho$ and $z$, respectively, in the above limit. Equation~\eqref{eq:LClimit} is the LC metric with $\sigma=1=k$~\cite{Griffiths:2009dfa}. If we only switch off the angular momentum of the seed, we recover the Schwarzschild--LC spacetime \cite{Akbar:2015qna,Mazharimousavi:2024hrg,Barrientos:2024uuq}. In both cases, we end up with \emph{static} solutions to Einstein's field equations.  

\begin{figure}
        \centering
        \begin{tikzpicture}
        \begin{groupplot}[group style={group size=2 by 2, horizontal sep=0.51cm, vertical sep=0.51cm}]
        \nextgroupplot[%
        enlargelimits=false,
        axis on top,
        xlabel= $\mathrm{x}$,
        ylabel= $\mathrm{z}$,
        xlabel near ticks,
        ylabel near ticks,
        xticklabel pos=top,
        yticklabel pos=left,
        scaled y ticks=false,
        scaled x ticks=false,
        minor x tick num=10,
        minor y tick num=10,
        ytick distance = {1},
        xtick distance = {1}
        ]
        \addplot graphics
            [xmin=-4,xmax=4,ymin=-4,ymax=4]
                {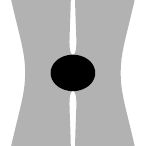};
        \node[anchor= center] at (rel axis cs:.1,.9) 
        {(a)};
        \nextgroupplot[%
        enlargelimits=false,
        axis on top,
        xlabel= $\mathrm{x}$,
        ylabel= $\mathrm{z}$,
        xlabel near ticks,
        ylabel near ticks,
        xticklabel pos=top,
        yticklabel pos=right,
        scaled y ticks=false,
        scaled x ticks=false,
        minor x tick num=10,
        minor y tick num=10,
        ytick distance = {1},
        xtick distance = {1}
        ]
        \addplot graphics
            [xmin=-4,xmax=4,ymin=-4,ymax=4]
                {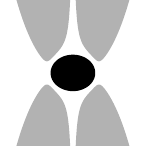};
        \node[anchor= center] at (rel axis cs:0.1,.9) 
        {(b)};
        \nextgroupplot[%
        enlargelimits=false,
        axis on top,
        xlabel= $\mathrm{x}$,
        ylabel= $\mathrm{z}$,
        xlabel near ticks,
        ylabel near ticks,
        xticklabel pos=bottom,
        yticklabel pos=left,
        scaled y ticks=false,
        scaled x ticks=false,
        minor x tick num=10,
        minor y tick num=10,
        ytick distance = {1},
        xtick distance = {1}
        ]
        \addplot graphics
            [xmin=-4,xmax=4,ymin=-4,ymax=4]
                {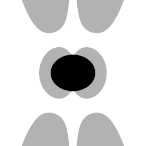};
        \node[anchor= center] at (rel axis cs:0.1,.9) 
        {(c)};
        \nextgroupplot[%
        enlargelimits=false,
        axis on top,
        xlabel= $\mathrm{x}$,
        ylabel= $\mathrm{z}$,
        xlabel near ticks,
        ylabel near ticks,
        xticklabel pos=bottom,
        yticklabel pos=right,
        scaled y ticks=false,
        scaled x ticks=false,
        minor x tick num=10,
        minor y tick num=10,
        ytick distance = {1},
        xtick distance = {1}
        ]
        \addplot graphics
            [xmin=-4,xmax=4,ymin=-4,ymax=4]
                {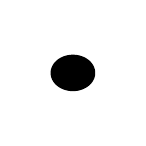};
        \addplot graphics
            [xmin=-4,xmax=4,ymin=-4,ymax=4]
                {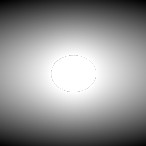};
        \node[anchor= center] at (rel axis cs:0.1,.9) 
        {\textcolor{white}{(d)}};
        \end{groupplot}
        \end{tikzpicture}
        \caption[]{Cross section, taken at $\mathrm{y}=0$, of the ergoregion (gray fill) dressing the event-horizon of a Kerr--LC black hole with $r_+=1$ and $r_-=1/2$ in rotating frames with angular velocity $\alpha=0$ (a), $\alpha=1.64$ (b), and $\alpha=3$ (c). In panel (d), we provide a density map of the absolute difference $\Delta\Omega$ in the above spacetime. We use total white for $\Delta\Omega=0$. The darker the color is, the bigger the difference becomes. The black ellipsoid represents the black hole.}
        \label{fig:ErgoPlots}
        \end{figure}

Now, let us scrutinize the ergoregions in the new spacetime. We defined them as the regions where $\bm{\partial}_{t^{(\alpha)}}$ becomes spacelike. Thus, they must obey the inequality
\begin{equation}
    256 a^2 m^2 f^2 \Omega^{(\alpha)2} - \rho^2\geq 0,\label{eq:ErgoDefIneq}
\end{equation}
which we visualize in \autoref{fig:ErgoPlots} using oblate spheroidal coordinates 
\begin{equation}
    \begin{split}
        \mathrm{x}&=\sqrt{(r^2+a^2)(1-x^2)}\cos\varphi,\\
        \mathrm{y}&=\sqrt{(r^2+a^2)(1-x^2)}\sin\varphi,\quad \mathrm{z}=rx.
    \end{split}
\end{equation}
In that figure, we observe cross sections with dramatically different shapes due to the dependence of~\eqref{eq:ErgoDefIneq} on the frame via $\Omega^{(\alpha)}$. If we define static observers as those whose worldlines are integral curves of $\bm{\partial}_{t^{(\alpha)}}$, then the findings in different rotating frames can even be contradictory. For example, in the rotating frame of panel (b), we can have a static observer ``hovering'' infinitesimally close to the black hole, which is impossible in the frames of panels (a) and (c). In the frame of panel (a), there is a radius at which no observer can be at rest at any altitude. That is not the case in the frames of panels (b) and (c). Interestingly, in the frame of panel (b), the ergoregions resemble those in the electromagnetic swirling Schwarzschild spacetime~\cite{Barrientos:2024pkt}. For completeness, we also include a density plot of $\Delta\Omega$ in panel (d). 

Nevertheless, we can exploit the fact that the asymptotic geometry is frame-independent to extract a universal result. Since $\Omega^{(\alpha)}$ grows as $\Omega^{(0)}$ in all directions at infinity, finding a direction where~\eqref{eq:ErgoDefIneq} holds asymptotically proves our initial claim: that there is no asymptotically timelike Killing vector everywhere, or, equivalently, that the ergoregions extend to infinity in the Kerr--LC spacetime. If we redefine $r=\hat{r}\epsilon$ and $x=(\hat{x}/\epsilon^2)-1$, take the limit $\epsilon\to\infty$, and set $\epsilon=1$ afterwards, we find that 
\begin{equation}
    g_{tt}\sim \frac{128a^2m^2\hat{x}\hat{r}^4}{4a^2m^2+\hat{x}^2\hat{r}^4}>0,
\end{equation}
which proves the claim indeed. We should mention that $|\bm{\partial}_{t^{(\alpha)}}|^2$ still blows up at $r=0=x$. However, since we showed that curvature invariants are regular there, the above metric singularity is just a coordinate artefact.  

Finally, let us close this section by finding the Kerr--Schild representation of the Kerr--LC spacetime, if it exists. It is well-established that the Kerr metric can be cast into the Kerr--Schild form, where the seed metric is Minkowski in special spheroidal coordinates (see, for example, Ref.~\cite{Ayon-Beato:2015nvz}). One may then naturally ask if the Kerr--LC metric also admits such a representation, namely
\begin{equation}
    \diff s^2 = \diff s_0^2 + H(r,x)\,\bm{l}\otimes \bm{l},
\end{equation}
where $\bm{l}$ generates the null geodesics,\footnote{Need not be shear-free, since the spacetime is algebraically general.} and $\diff s_0^2$ denotes the seed metric~\eqref{eq:LClimit} allowing the Kerr--Schild representation of our spacetime. For the moment, let us consider a general 1-form $\bm{l}=\diff t + A \diff r + B \diff x + C\diff \phi$, with $A,\ B,$ and $C$ being functions of $r$ and $x$. The arbitrary function that should be multiplying $\diff t$ can always be set to one by redefining $H$. However, there is a key difference with Kerr: the component $l^x$ cannot be trivial, because $g_{xx}$ depends explicitly on the mass parameter $m$. On the other hand, since the Kerr--LC metric is a vacuum solution, the circularity theorem applies, and one should be able to perform coordinate transformations like $t \to t+\alpha(r,x)$ and $\phi \to \phi + \beta(r,x)$ to eliminate the terms that break circularity. Consequently, the metric components along the Killing directions, i.e., $g_{tt}$, $g_{\phi\phi}$, and $g_{t\phi}$, must be entirely determined by the functions $H$ and $C$. That overdetermined system does not admit a solution, unlike in Kerr. This is a particularly interesting result, as the static Schwarzschild--LC metric admits a Kerr--Schild representation~\cite{Barrientos:2024uuq}. 
\vspace{-0.61cm}
\section{Conclusions: Novelty of the Kerr--LC solution and further perspectives}\label{secIV}
This letter has been devoted to the construction of a novel rotating vacuum solution of Einstein's field equations, the Kerr--Levi-Civita spacetime. The new geometry is a rotating generalization of the recently studied Schwarzschild--Levi-Civita black hole \cite{Akbar:2015qna,Mazharimousavi:2024hrg,Barrientos:2024uuq}. The solution was generated from the Kerr metric by exploiting a discrete symmetry of the field equations in the complex potential formalism, known as \emph{inversion}.

The Kerr--LC spacetime has a remarkable amount of good properties that make it attractive: it is free of curvature singularities, conical defects, spinning strings, and closed timelike curves. When $m^2>a^2$, it describes a rotating black hole with the same horizon radii as the Kerr black hole, but with the above nice features, which also hold in the extremal and super-spinning regimes. It appears that the interplay between the rotation of the Kerr seed and that of a generalized Levi-Civita asymptotic geometry regularizes the ring singularity of the Kerr black hole. As a matter of fact, that ``softening'' of the interior also happens in the Kerr--Swirling spacetime, studied in~\cite{Astorino:2022prj}, where the authors erroneously reported that the ring singularity remains. Here, we achieve that curvature regularization without introducing additional parameters, recovering the Schwarzschild--LC spacetime in the static limit.  

We claimed that in all spacetimes with a similar asymptotic geometry it is impossible to define a Killing vector that is asymptotically timelike everywhere. Since the ergo-geometry depends on the choice of frame and no preferred frame exists, we cannot really distinguish between the three qualitatively different profiles in \autoref{fig:ErgoPlots}. The frame-independent conclusions are that the ergoregion extends to infinity in some directions and that ZAMOs experience a more intense dragging as they move farther from the source. 

A natural question arises regarding the effect of inversion in the case of a Kerr seed, expressed in the electric WLP form. In that case, the transformed Ernst potential remains asymptotically flat. By virtue of the uniqueness of the Kerr family \cite{Carter:1971zc,Israel:1967za}, the transformed metric must again be the Kerr metric. In fact, the ``electric'' inversion does not truly transform any solution in the Pleba\'nski--Demia\'nski class \cite{Plebanski:1976gy}, and the importance of expressing the seed metric in a magnetic form becomes clear.

Let us now address the intrinsic novelty of the Kerr--LC geometry. One may naively believe that the Kerr--LC class of solutions is contained in the Kerr--Swirling family~\cite{DiPinto:2025yaa}, speculating that the former should follow from the latter when the Ehlers parameter approaches an appropriate limit and/or via diffeomorphisms and parameter redefinitions. That guesswork is based on the fact that the inversion operator can be expressed as the composition  
\begin{equation}
    \mathcal{D}_{\iu/b}\circ E_{1/b}\circ G^{g}_{b}
\end{equation}
in the limit $b\to\infty$, where 
\begin{subequations}
    \begin{align}
        \mathcal{D}_\alpha&:\alpha (\alpha^*\mathcal{E},\Phi),\\
        E_j&:(\mathcal{E},\Phi)\mapsto(\mathcal{E},\Phi)(1+\iu j \mathcal{E})^{-1},\\
        G^g_b&:(\mathcal{E}+\iu b,\Phi),
    \end{align}
\end{subequations}
are the rescaling-duality, Ehlers, and gravitational gauge transformations, respectively.\footnote{The parameters $j$ and $b$ are real, while the parameter $\alpha$ is complex.} However, there are various subtle points that \emph{falsify} the above speculation. 

First, the rescaling-duality transformation is a non-trivial transformation that, \emph{in general}, produces a new solution which cannot be mapped to other solutions via coordinate reparametrizations and parameter redefinitions (some exceptions may exist). Second, gravitational gauge transformations encode the freedom to shift the gravitational twisted potential $\chi$ by a constant. That freedom exists only in potential space and does not affect the physical metric somehow, since the differential equation for $\chi$ involves only its gradient, meaning that $\chi$ and $\chi+c$ result in the same $\omega$. Of course, $\omega$ is always defined up to a constant since only its gradient appears in the defining equation. Since $\omega$ appears in the metric, the freedom in the definition of $\omega$ translates to a one-parameter family of solutions, where the parameter, say $b$, is the angular speed of a rigidly rotating reference frame. Therefore, {the gravitational gauge transformation and the freedom in the definition of $\omega$ are two completely unrelated things}. Moreover, as we will show, the stated freedom to shift $\chi$ is not so innocent if followed by an Ehlers or a Harrison transformation, for it actually manifests itself in the spacetime metric in those cases. In what follows, we will demonstrate that the Kerr--Swirling metric follows from a more general metric when we fix the trivial gravitational gauge, whereas the Kerr--LC metric follows from that very same general metric for another choice of gravitational gauge, followed by a rescaling transformation which removes the Ehlers parameter.   

For starters, let us state that inversion can be written as
\begin{equation}
    G^g_{j}\circ\mathcal{D}_{\iu j} \circ E_j\circ G^g_{1/j}\label{eq:InvAlt}
\end{equation}
at the level of symmetry operators. As promised, the above proves that no limit whatsoever is necessary in the process. Do also note that~\eqref{eq:InvAlt} and $\mathcal{D}_{\iu j} \circ E_j\circ G^g_{1/j}$ produce the same solution, since the last gravitational gauge transformation cannot affect the form of the physical metric. Now, the more general metric that we previously talked about can be found by acting on the potentials of the Kerr seed (in magnetic WLP form) with the ``generalized'' Ehlers transformation operator $E_j\circ G^g_a$, which encodes the freedom in the definition of the gravitational twisted potential. The result is the metric~\eqref{mWLPspherical} with $f_0$ and $\omega_0$ replaced by 
    \begin{subequations}
        \begin{align}
            \hat{f}(r,x) &= \frac{f_0}{j^2f_0^2+(1-bj+j\chi_0)^2},\\
            \hat{\omega}(r,x) &=\frac{\Delta_x}{f_0\varrho^2}(2amr-j h_{(1)}+j^2 h_{(2)}),
        \end{align}\label{eq:GSol}
    \end{subequations}
\begin{widetext}
respectively, where 
    \begin{subequations}
        \begin{align}
            h_{(1)}(r,x)&=4abmr+4(r^3-ma^2)\Delta_r x + 4 a^2(r-m)\Delta_r x^3,\\
            h_{(2)}(r,x)&=2 a m [2 a^2 r^2 (r+3 m)-a^4(r+2 m)+r (b^2+3 r^4)] + 4b(r^3-ma^2)\Delta_r x+12 am\Delta_r[r^3+a^2(r-m)]x^2\nonumber\\
            &\hphantom{=}+4 a^2b(r-m)\Delta_r x^3-2am[r^3-a^2(3r-2m)]\Delta_r x^4.
        \end{align}
    \end{subequations}
\end{widetext}

The above solution assumes the form of the Kerr--Swirling solution when we choose the trivial gravitational gauge, i.e., $b=0$. On the other hand, if we choose the gauge $b=1/j$, we get a metric with
\begin{equation}
    \hat{f} = f/j^2,\quad \hat{\omega}=j^2\omega.\label{eq:RescaledLC}
\end{equation}
There is absolutely no diffeomorphism/parameter redefinition/limit that maps the above into the Kerr--Swirling solution; the two metrics correspond, simply put, to the general solution~\eqref{eq:GSol} in different gravitational gauges! Similarly, there is no physical-space invariance that casts the metric with functions~\eqref{eq:RescaledLC} into the Kerr--LC metric.\footnote{At first sight, one may think that a global Weyl rescaling may do the job, but this is ultimately a fallacy.} Indeed, the only way to move from the former to the latter is to further apply a rescaling-duality transformation with parameter $\iu j$, which completely removes the Ehlers parameter from the solution (recall that inversion is a discrete symmetry). Hence, the Kerr--LC solution is truly a novel, rotating solution of GR in vacuum. 

Lastly, let us remark that the difference between inversion and an Ehlers transformation is most obvious in the case of static seeds, where inversion produces a static target solution, while a simple Ehlers transformation generates a truly stationary target solution, i.e., there is no boost that maps one to the other. 

Consequently, the Kerr--Levi-Civita spacetime proves to be a highly appealing, regular, rotating vacuum solution in Einstein's theory, opening up several research directions to pursue. Among these, a thorough analysis of geodesic motion is of particular importance. Exploring hidden symmetries generated by higher-order Killing tensor fields would also enhance our understanding of the spacetime. Moreover, one should study black hole shadows, gravitational lensing, and related phenomenology to assess the astrophysical importance of this novel solution. Finally, the computation of conserved charges is equally relevant.
\acknowledgments
The work of J.B. is supported by FONDECYT Postdoctorado grant 3230596.  A.C. is partially supported by FONDECYT grant 1250318 and by the GA{\v C}R 22-14791S grant from the Czech Science Foundation. M.H. gratefully acknowledges the University of Paris-Saclay for its warm hospitality during the development of this project. The work of K.M. is funded by Beca Nacional de Doctorado ANID grant No. 21231943.
\appendix
\section{Regularity of curvature invariants}\label{appA}
In this appendix section, we show that there is no curvature singularity in the Kerr--LC spacetime. We start with the Kretschmann scalar, $\mathcal{K}=R^{\mu\nu}{}_{\rho\sigma}R^{\rho\sigma}{}_{\mu\nu}$, which reads 
\begin{equation}    \mathcal{K}=\frac{\mathcal{A}(r,x)}{\varrho^{12}\left(f_0^2+\chi_0^2\right)^6},
\end{equation}
where $\mathcal{A}$ is an involved and everywhere non-trivial polynomial in $r$ and $x$ that we need not present here, and $f_0^2+\chi_0^2$ is another polynomial in the above variables, divided by $\varrho^2$, that is also non-vanishing everywhere, so that the denominator of $\mathcal{K}$ conveniently assumes the form of a polynomial that does not vanish at any $r$ and $x$. Recalling that $\varrho^2=0$ at $r=0=x$, we see that inversion effectively works as a pole-removal tool. The Kretschmann scalar acquires the value $-9/(4a^8 m^4)$ at $r=0=x$. It reduces to the (singular) Kretschmann scalar in the LC and Schwarzschild--LC spacetimes, when we take the limits $m,J\to 0$ ($a$ finite) and $m\to 0$, respectively. 

Although it is extremely unlikely, the singularity could, in theory, appear in higher-order curvature invariants. To mathematically prove regularity, we must verify that the components $R^{AB}{}_{CD}$ of the Riemann tensor in an orthonormal basis are regular. We choose the frame field dual to the co-basis
\begin{equation}
  \begin{aligned}
\boldsymbol{\theta}^0&=\sqrt{\frac{\Delta_r\Delta_x}{f}}\diff t,\quad
    \boldsymbol{\theta}^1=\frac{\eu^\gamma}{\sqrt{f\Delta_r}}\diff r,\\
    \boldsymbol{\theta}^2&=-\frac{\eu^\gamma}{\sqrt{f\Delta_x}}\diff x,\quad
    \boldsymbol{\theta}^3= \sqrt{f}\left(16 a^2 m^2\diff\varphi - \omega\diff t\right),
  \end{aligned}
\end{equation}
and find out that some components ($R^{01}{}_{01}$ for example) have a pole at the ring singularity of the Kerr spacetime. Of course, that does not imply that there is a curvature singularity because the poles could disappear when contracting Riemann tensors to build invariants, as in the case of the regular Kretschmann scalar. In fact, we find that curvature invariants up to sixth-order are regular everywhere, their value at $r=0=x$ being proportional to $(a^4m^2)^{-k}$, where $k$ denotes the order of the invariant. Therefore, we have good reason to call the spacetime regular in the sense that ``strong'' singularities (infinite curvature invariants) are absent. We cannot exclude the presence of ``weak'' singularities in the wider Penrose sense of geodesic incompleteness.

\bibliography{apssamp}

\end{document}